\documentstyle{article}

\begin{document}

\begin{center}
{\large {\bf Reinterpretation of Matter-Wave Interference Experiments}}

{\large {\bf Based on the Local-Ether Wave Equation}}

$\ $

{\sf Ching-Chuan Su}

Department of Electrical Engineering

National Tsinghua University

Hsinchu, Taiwan
\end{center}

$\ $

\noindent {\bf Abstract }-- Based on the local-ether wave equation for free
particle, the dispersion of matter wave is examined. From the dispersion
relation, the angular frequency and wavelength of matter wave are derived.
These formulas look like the postulates of de Broglie in conjunction with
the Lorentz mass-variation law. However, the fundamental difference is that
for terrestrial particles their speeds are referred specifically to a
geocentric inertial frame and hence incorporate the speed due to earth's
rotation. Thus the local-ether model predicts an east-west directional
anisotropy both in mass and wavelength. Meanwhile, in spite of the
restriction on reference frame, the local-ether model can account for the
matter-wave interference experiments of the Bragg reflection and the Sagnac
effect. For electron wave, the effects of earth's rotation are negligible
and the derived Bragg angle is actually in accord with the Davisson-Germer
experiment, as examined within the present precision. On the other hand, the
local-ether model leads to a directional anisotropy in the Bragg angle in
neutron diffraction. The predicted anisotropy due to earth's rotation then
provide a means to test the local-ether wave equation.

$\ $

$\ $

\noindent {\large {\bf 1. Introduction}}

The wave nature of particle has been initiated by de Broglie in 1924 by
postulating that a particle is associated with a matter wave of which the
angular frequency and the wavelength are related to the energy and the
momentum of the particle, respectively [1]. The energy and momentum in turn
are associated with the speed-dependent mass which was first introduced by
Lorentz in 1904 [2]. Shortly, the matter wavelength has been demonstrated in
the Bragg reflection of electron beam from a crystal by Davisson and Germer
in 1927 [1]. Furthermore, based on the phase difference between two coherent
beams of electron, neutron, or atom, various matter-wave interference
experiments have been reported to demonstrate the Bragg reflection, the
double-slit diffraction, the gravitational effect, and the Sagnac effect.

Recently, we have presented the local-ether model of wave propagation [3].
That is, electromagnetic wave can be viewed as to propagate via a medium
like the ether. However, the ether is not universal. It is supposed that in
the region under sufficient influence of the gravity due to the Earth, the
Sun, or another celestial body, there forms a local ether which in turn
moves with the gravitational potential of the respective body. For
earthbound waves, the medium is the earth local ether which as well as
earth's gravitational potential is stationary in an ECI (earth-centered
inertial) frame, while the sun local ether for interplanetary waves is
stationary in a heliocentric inertial frame. It has been shown elaborately
that the propagation of earthbound electromagnetic waves is referred
specifically to an ECI frame, as demonstrated in the high-precision
experiments of GPS (global positioning system), the intercontinental
microwave link, and of the Sagnac loop interferometry [3].

Further, matter wave is supposed to follow the local-ether model. Thereby,
we have presented a wave equation for a particle of charge $q$ and natural
frequency $\omega _{0}$ in the presence of the gravitational and the
electrical scalar potentials [4-6]. Under the ordinary condition of low
particle speed, the local-ether wave equation has been shown to lead to a
unified quantum theory of gravitational and electromagnetic forces [5, 6].
Furthermore, it is found that the gravitational mass associated with the
gravitational force and the inertial mass under the influence of the
electromagnetic force are identical to the natural frequency, aside from a
common scaling factor. Thereby, the local-ether wave equation leads to the
important consequence of the identity of gravitational and inertial mass and
their physical origin. When the restriction on the particle speed is
removed, the local-ether wave equation leads to the east-west directional
anisotropy in mass, quantum state energy, and hence in clock rate which in
turn has been demonstrated in the Hafele-Keating experiment with
circumnavigation atomic clocks [4].

In this investigation, we explore more consequences of the local-ether wave
equation{\rm .} From the wave equation for a free particle, the dispersion
of matter wave is derived. Then, from the dispersion relation, the angular
frequency and wavelength of matter wave and the associated speed-dependent
mass of particle are derived. According to the local-ether model, the
velocity of earthbound particles that determines the mass and the wave
properties is referred specifically to an ECI frame and hence incorporates
the linear velocity due to earth's rotation. Accordingly, the matter-wave
interference experiments of the Bragg reflection, the double-slit
diffraction, and of the Sagnac effect are reexamined, particularly the
effects of earth's rotational and orbital motions.

$\ $

\noindent {\large {\bf 2. Local-Ether Wave Equation for Free Particle}}

It is supposed that a free particle is represented by a wavefunction $\Psi $
which in turn is governed by the nonhomogeneous wave equation proposed to be 
$$
\left\{ \nabla ^{2}-\frac{1}{c^{2}}\frac{\partial ^{2}}{\partial t^{2}}%
\right\} \Psi ({\bf r},t)=\frac{\omega _{0}^{2}}{c^{2}}\Psi ({\bf r},t),%
\eqno
(1) 
$$
where $c$ is the speed of light and $\omega _{0}$ is the natural frequency
associated with the particle. This equation is a simplified form of the
local-ether wave equation presented in [4-6] by dropping the gravitational
and the electric scalar potentials. For the case where the natural frequency
is zero, the equation reduces to the homogeneous wave equation governing
electromagnetic wave in free space. Further, by replacing $c$ with another
speed, the equation reduces to the wave equation governing a mechanical wave.

The natural frequency is a constant and has been found to be the origin of
the rest mass of the particle associated with wavefunction $\Psi $. It has
been shown from the local-ether wave equation under the gravitational and
the electric potential potentials that the natural frequency $\omega _{0}$
is related to the gravitational and the inertial mass $m_{0}$ of a low-speed
particle by the familiar form of [4-6] 
$$
m_{0}=\frac{\hbar }{c^{2}}\omega _{0},\eqno
(2) 
$$
where $\hbar $ is Planck's constant divided by $2\pi $. This relation states
that the rest mass of a particle is just its natural frequency, aside from
the scaling factor $\hbar /c^{2}$.

Since matter wave is supposed to follow the local-ether model, the time
derivative in the proposed wave equation is referred specifically to an ECI
frame for earthbound particles. As the time derivative is referred to a
particular frame, its value remains unchanged when observed in different
reference frames. This feature is identical to that in a mechanical wave
motion. The proposed wave equation looks like the free-space Klein-Gordan
equation [7, 8] and like the equation dealing with spin given in [9], except
the fundamental difference in the reference frame of the time derivative.

From the wave equation, it is seen that the wavefunction $\Psi $ oscillates
at the natural frequency when its spatial variations vanish. As the
wavefunction starts to vary spatially, its temporal variation will increase
accordingly. The spatial and the temporal variations of a wavefunction are
important characteristics of wave motion and are expected to play an
important role in determining the physical properties of the particle
associated with the wavefunction.

$\ $

\noindent {\large {\bf 3. Dispersion of Matter Wave}}

Suppose that the wavefunction $\Psi $ is a wave packet composed of plane
waves of a narrow bandwidth. Each component of the plane waves is of the
form $\psi _{0}e^{ikx}e^{-i\omega t}$, where $\omega $ is the angular
frequency, $k$ the propagation constant, and $\psi _{0}$ an arbitrary
constant. Then, for each of the plane waves, the wave equation leads to an
algebraic equation 
$$
\omega ^{2}-c^{2}k^{2}=\omega _{0}^{2}.\eqno
(3) 
$$
It is seen that the angular frequency and the propagation constant are
related to each other. Further, owing to the presence of the natural
frequency $\omega _{0}$, the relation between $\omega $ and $k$ becomes
nonlinear and hence the matter wave is dispersive. For electromagnetic wave
with a zero natural frequency, the dispersion then vanishes. On the other
hand, the dispersion relation is important in determining the properties of
matter wave.

It is known that the peak of a wave packet moves at its group velocity. Thus
the speed $v$ of a particle can be given by the group speed $v_{g}$ of the
associated wave packet, that is, $v=v_{g}=d\omega /dk$. Then, from the
preceding dispersion relation, one has $\omega d\omega =c^{2}kdk$ which in
turn yields the propagation constant-speed relation 
$$
k=\frac{\omega }{c^{2}}v.\eqno
(4) 
$$
It is noted that the propagation constant $k$ is proportional to the speed $%
v $ with the angular frequency $\omega $ as a ratio.

On substituting this relation back into the dispersion relation, it is seen
that the frequency $\omega $ in turn depends on the speed $v$ as 
$$
\omega =\frac{\omega _{0}}{\sqrt{1-v^{2}/c^{2}}}.\eqno
(5) 
$$
In terms of the natural frequency $\omega _{0}$, the propagation
constant-speed relation becomes 
$$
k=\frac{\omega _{0}}{c^{2}}\frac{1}{\sqrt{1-v^{2}/c^{2}}}v{\bf .}\eqno
(6) 
$$
Since wavelength is inversely proportional to propagation constant, the
wavelength of a harmonic-like matter wave packet also depends on the
particle speed.

In terms of the rest mass $m_{0}$, the angular frequency and the propagation
constant take the form of 
$$
\hbar \omega =\frac{m_{0}c^{2}}{\sqrt{1-v^{2}/c^{2}}}\eqno
(7) 
$$
and 
$$
\hbar k=\frac{m_{0}v}{\sqrt{1-v^{2}/c^{2}}},\eqno
(8) 
$$
respectively. It is of essence to note that the preceding two formulas look
like the postulates of de Broglie. However, the fundamental difference is
that the particle speed here is referred specifically to the local-ether
frame, which is an ECI frame for earthbound particles. Ignoring this
difference in reference frame, the famous postulates of de Broglie are then
consequences derived from the dispersion of matter wave.

From the frequency-speed relation (5), one immediately has 
$$
v=c\sqrt{1-\omega _{0}^{2}/\omega ^{2}}.\eqno
(9) 
$$
It is seen that the particle speed $v$ increases monotonically with the
angular frequency $\omega $. Further, the speed of a free particle is then
limited to $c$, as referred to the local-ether frame. For this limiting
case, the propagation constant $k$ and hence the angular frequency $\omega $
are large enough, such that the natural frequency $\omega _{0}$ in the
dispersion relation (3) can be neglected. As a result, the propagation
characteristics of such a matter wave are close to those of electromagnetic
wave. For terrestrial high-energy electrons emitted spontaneously from
radiative elements or accelerated in a synchrotron, the limiting speed is
expected to be referred to an ECI frame. However, as the linear speed due to
earth's rotation is relatively low, it makes no substantial difference in
measurement if the speed is referred instead to a geostationary laboratory
frame, as done tacitly in common practice.

A dispersion relation quite similar to (3) can be found in the plane wave
propagating in a plasma, with the natural frequency $\omega _{0}$ being
replaced with the plasma frequency [10]. Another similar dispersion can be
found in the guided mode propagating in a waveguide, with the natural
frequency $\omega _{0}$ being replaced with the cutoff frequency [10]. Hence
a frequency-speed relation similar to (5) or (9) can be found in a plasma or
a waveguide.

$\ $

\noindent {\large {\bf 4. Mass, Energy, and\ Momentum}}

The formulas of angular frequency and propagation constant can be written in
a more compact way by introducing the speed-dependent mass $m$ defined in
terms of the rest mass $m_{0}$ in a familiar form of 
$$
m=m_{0}\frac{1}{\sqrt{1-v^{2}/c^{2}}}.\eqno
(10) 
$$
Then we have the frequency-mass relation 
$$
\hbar \omega =mc^{2}\eqno
(11) 
$$
and the propagation vector-velocity relation 
$$
\hbar {\bf k}=m{\bf v}.\eqno
(12) 
$$
This speed-dependent mass looks like the famous Lorentz mass-variation law.
However, the fundamental difference is that the particle speed $v$ in (10)
is referred specifically to the local-ether frame, rather than a laboratory
frame or any other. Thus the speeds of terrestrial particles incorporate the
one due to earth's rotation, while the corresponding wave properties are
entirely independent of earth's orbital motion around the Sun or others.

From the preceding relations, it is seen that the rest mass $m_{0}$ and the
speed-dependent mass $m$ of a particle are just the natural frequency and
the angular frequency of the associated matter wave, respectively, aside
from the common scaling factor $\hbar /c^{2}$. Thereby, based on the
local-ether wave equation, the mass originates from the temporal variation
of matter wave. And the mass variation originates from the dispersion of
matter wave, which in turn is due to the natural frequency. On the other
hand, it is noticed that the standard derivation of the mass variation is
through a quite indirect way by dealing with a case associated with an
elastic or inelastic collision between two identical particles [11-13].

For a high-speed terrestrial particle, it may make no substantial difference
whether the particle speed is referred to an ECI frame or to the ground.
Nevertheless, one consequence of the local-ether model is that the
speed-dependent mass $m$ as well as the matter wavelength is expected to
possess an east-west directional anisotropy due to earth's rotation. That
is, for a given ground speed, the earthbound particles moving eastward have
the highest speed with respect to an ECI frame and hence have the heaviest
mass, while those moving westward, the lightest. This anisotropy in mass in
conjunction with the mass-dependence of quantum energy of the matter wave
bounded in atoms has been adopted to account for the east-west directional
anisotropy in the clock rate observed in the Hafele-Keating experiment with
circumnavigation atomic clocks [4].

Similar to those in quantum mechanics, the {\it momentum} and the {\it energy%
} of a particle are defined as the expectation values of the spatial and the
temporal derivatives, $\left\langle -i\hbar \nabla \right\rangle $ and $%
\left\langle i\hbar \partial /\partial t\right\rangle $, respectively, where
the expectation value of an operator $O$ is evaluated in terms of the
wavefunction $\Psi $ as $\left\langle O\right\rangle =\int \Psi ^{*}O\Psi d%
{\bf r}$. Thus, for a space-harmonic wave, the momentum is given by ${\bf p}%
=\hbar {\bf k}$. Then, according to the propagation vector-velocity relation
(12), the momentum becomes a familiar form of ${\bf p}=m{\bf v}$. However,
the particle velocity ${\bf v}$ is referred specifically to the local-ether
frame. From the local-ether wave equation, a first-order time evolution
equation, similar to Schr\"{o}dinger's equation, has been derived. From this
equation it has been shown that the velocity of a particle defined as the
time derivative of expectation value of the position vector corresponds to
the expectation value of a spatial derivative as ${\bf v}=\left\langle
-i\hbar \nabla \right\rangle /m_{0}$, where the velocity is referred to the
local-ether frame and the corresponding speed is supposed to be much lower
than $c$ [4-6]. It is of interesting to note that for a wave packet, the
particle speed derived from the evolution equation is identical to the
aforementioned group velocity derived from the dispersion relation.

Furthermore, for a time-harmonic wave, the energy is given by ${\cal E}%
=\hbar \omega $. Then, according to the frequency-mass relation (11), the
energy becomes a familiar form of 
$$
{\cal E}=mc^{2}.\eqno
(13) 
$$
Ignoring the difference in reference frame of the particle speed, this is
just the famous energy-mass relation first introduced by Poincar\'{e} in
1900 [14]. For a low-speed particle, the mass-variation law (10) leads to 
$$
\hbar \omega -\hbar \omega _{0}=mc^{2}-m_{0}c^{2}=\frac{1}{2}m_{0}v^{2}.%
\eqno
(14) 
$$
Thus, due to the dispersion of matter wave, the variation in the angular
frequency or in the mass corresponds to the {\it kinetic energy} for a
low-speed particle. When the spatial variation of $\Psi $ is weak, the
temporal variation of $\Psi $ is close to the harmonic $e^{-i\omega _{0}t}$
and then the wavefunction can be given as $\Psi ({\bf r},t)=\psi ({\bf r}%
,t)e^{-i\omega _{0}t}$, where the temporal variation of the wavefunction $%
\psi $ is weak. The aforementioned evolution equation derived from the wave
equation is expressed in terms of this reduced wavefunction $\psi $. It is
seen that the energy and the kinetic energy are associated with the
expectation values of the time derivative $i\hbar \partial /\partial t$
evaluated in terms of wavefunctions $\Psi $ and $\psi $, respectively.

As the particle velocity is referred to a specific frame, all the values of
the angular frequency, the propagation vector, the energy, the kinetic
energy, and of the momentum remain unchanged when observed in different
reference frames. For earthbound particles, the momentum and the kinetic
energy are then referred specifically to an ECI frame. This differs from the
common understanding, since the conventional momentum and kinetic energy are
not attached to a specific frame and hence are different in different
frames. However, in what follows we show that the local-ether momentum and
kinetic energy indeed comply with the conservation laws of the conventional
momentum and kinetic energy.

Consider the collision between two low-speed particles of rest masses $m_{1}$
and $m_{2}$. It is supposed that the sum of the propagation constants and
the one of the angular frequencies of the two particles remain fixed during
the collision. Thus 
$$
m_{1}{\bf v}_{1}+m_{2}{\bf v}_{2}=m_{1}{\bf v}_{3}+m_{2}{\bf v}_{4}\eqno
(15)
$$
and 
$$
m_{1}v_{1}^{2}+m_{2}v_{2}^{2}=m_{1}v_{3}^{2}+m_{2}v_{4}^{2},\eqno
(16)
$$
where ${\bf v}_{1}$ and ${\bf v}_{3}$ are the velocities of the particle of $%
m_{1}$ before and after the collision, respectively, ${\bf v}_{2}$ and ${\bf %
v}_{4}$ are those of the particle of $m_{2}$, and all the velocities are
referred to the local-ether frame.

Then, for an arbitrary velocity ${\bf v}_{0}$, one immediately has 
$$
m_{1}({\bf v}_{1}-{\bf v}_{0})+m_{2}({\bf v}_{2}-{\bf v}_{0})=m_{1}({\bf v}%
_{3}-{\bf v}_{0})+m_{2}({\bf v}_{4}-{\bf v}_{0}).\eqno
(17) 
$$
This relation can be interpreted in the way that based on Galilean
transformations the conventional momentum conserves as observed in a
laboratory frame moving at the velocity ${\bf v}_{0}$ with respect to the
local-ether frame. Further, by a direct expansion and by a use of (15), one
has 
$$
m_{1}({\bf v}_{1}-{\bf v}_{0})^{2}+m_{2}({\bf v}_{2}-{\bf v}_{0})^{2}=m_{1}(%
{\bf v}_{3}-{\bf v}_{0})^{2}+m_{2}({\bf v}_{4}-{\bf v}_{0})^{2}.\eqno
(18) 
$$
Therefore, it is seen that the conventional momentum and kinetic energy
conserve in any laboratory frame, in spite of the situation that their
values of respective particles are different in different frames. That is,
the conservation laws of conventional momentum and kinetic energy are
invariant under Galilean transformations and have no preferred frame, as
expected in classical mechanics. Thus the local-ether model leads to
consequences in accord with the conservation laws in classical mechanics and
with Galilean relativity.

For the case with $m_{2}\gg m_{1}$ and ${\bf v}_{2}={\bf v}_{0}$, it is seen
that ${\bf v}_{4}\simeq {\bf v}_{0}$ and then $({\bf v}_{3}-{\bf v}%
_{0})^{2}=({\bf v}_{1}-{\bf v}_{0})^{2}$. Thus the particle speed, the
magnitude of momentum, and the kinetic energy of a particle remain
substantially unchanged before and after its collision with a rigid plane,
when observed in the laboratory frame with respect to which the reflecting
plane is stationary. Meanwhile, these two particle speeds in general are
different when referred to the local-ether frame. Thus the wavelength of the
reflected matter wave is different from that of the incident one. Since the
difference between ${\bf v}_{3}$ and ${\bf v}_{1}$ is parallel to the normal 
$\hat{n}$ of the reflecting plane, it can be shown that 
$$
{\bf v}_{3}={\bf v}_{1}-2({\bf v}_{1}-{\bf v}_{0})\cdot \hat{n}\hat{n}.%
\eqno
(19) 
$$
Moreover, one has $\hat{n}{\bf \cdot }({\bf v}_{3}-{\bf v}_{0})=-\hat{n}{\bf %
\cdot }({\bf v}_{1}-{\bf v}_{0})$. When the reflecting point is fixed on a
rigid plane, the directions of the beams formed by the incident and the
reflected particles are determined by the particle velocities with respect
to the plane. Then the preceding relations lead to that the angle of
reflection of a particle beam from the plane is equal to the angle of
incidence, as observed in the laboratory frame. This consequence is simply
in accord with the famous Snell's law of reflection.

$\ $

\noindent {\large {\bf 5. Reexamination of Matter-Wave Interference
Experiments}}

In this section, we discuss the matter-wave interferometry, where two matter
waves are coherently split from a particle beam, propagate respectively
along two separate paths, and then are combined to cause an interference
depending on the phase difference. It is supposed that the phase variation $%
\phi $ of a matter wave over a differential path of directed length $d{\bf l}
$ along the particle beam is given by 
$$
\phi ={\bf k}\cdot d{\bf l,}\eqno
(20) 
$$
where the particle velocity ${\bf v}$ associated with the propagation vector 
${\bf k}$ ($=m{\bf v}/\hbar $) is referred specifically to the local-ether
frame, while the directed length $d{\bf l}$ is invariant in different
reference frames. This expression is identical to the one adopted in [15,
16], except the reference frame of the particle velocity ${\bf v}$. Thereby,
we reexamine the matter-wave interference experiments reported in the
literature demonstrating the Bragg reflection, the double-slit diffraction,
and the Sagnac effect, particularly the reference frame of particle velocity
and the effects of earth's motions.

$\ $

\noindent {\bf 5.1. Bragg reflection and Young's slit diffraction of matter
wave}

The Bragg reflection from a crystal is due to the constructive interference
among various reflected waves from successive lattice planes in parallel.
Analogous to the Bragg reflection of x-ray from a crystal, the wavelength
and the propagation constant of a matter wave can be determined from the
scattering from the surface of lattice planes of a known spacing by
measuring the Bragg angle $\theta _{B}$, the reflection angle corresponding
to the constructive interference.

For a crystal of lattice spacing $d$, the path-length difference between two
waves either incident upon or reflected from two consecutive lattice planes
is $d\cos \theta ^{\prime }$, where $\theta ^{\prime }$ is the angle of
incidence and reflection measured from the normal $\hat{n}$ of the
reflecting lattice plane when observed in the laboratory frame in which the
crystal is stationary. Then, according to the aforementioned Snell's law of
reflection, the Bragg condition of constructive reflection is given by 
$$
m(v_{p}+{\bf v}_{0}\cdot {\bf t})d\cos \theta ^{\prime }=n\pi \hbar ,\eqno
(21) 
$$
where ${\bf v}_{p}$ ($={\bf v}-{\bf v}_{0}$) represents the particle
velocity with respect to the laboratory frame which in turn moves at the
velocity ${\bf v}_{0}$ with respect to the local-ether frame, ${\bf t}$ is
one half of the sum of the unit vectors $\hat{v}_{p}$ representing the
directions of the incident and the reflected beams, and $n$ is a positive
integer. It is noted that the direction of ${\bf t}$ is parallel to the
lattice planes and its magnitude $t=\sin \theta ^{\prime }$. It is seen that
the Bragg angle depends on the laboratory speed and the orientation of the
lattice planes with respect to the laboratory velocity.

Consider the ordinary case where the particle speed $v$ is much lower than
the speed of light and hence the particle mass is substantially identical to
its rest mass $m_{0}$. In the mean time, the particle speed is supposed to
be high enough such that $v_{0}\ll v_{p}\ll c$. Thus the Bragg condition
reduces to 
$$
m_{0}v_{p}d\cos \theta ^{\prime }=n\pi \hbar .\eqno
(22)
$$
As the laboratory speed $v_{0}$ is omitted, this formula becomes invariant
under Galilean transformations. The preceding formula is identical to the
one given in [1], if the matter wavelength therein is understood to be
associated with the particle speed referred to the laboratory frame.

The matter wavelength has been demonstrated in the Davisson-Germer
experiment, where the scattering of accelerated electron beams from the
surface of a metal crystal was measured. Similarly, the matter wavelength
has also been demonstrated in the diffraction ring pattern of a high-speed
electron beam passing through a thin foil of polycrystalline gold [1]. More
direct evidence for matter wavelength can be provided by the diffraction
from macroscopic objects, such as Young's double slit. It is well known in
optics that the period in the interference fringe pattern is proportional to
the wavelength and to the inverse of the slit separation. Single-, double-,
or multiple-slit diffraction of matter wave has been demonstrated for
electron [17], neutron [18], and atom [19].

In these various experiments demonstrating the matter wavelength, the
reference frame of the particle velocity is not explicitly specified and is
supposed to be a geostationary laboratory frame. In what follows, we discuss
the issue of reference frame and the effects of earth's rotation. Consider
the isotropic case where the speed of the particles\ with respect to the
ground is independent of the beam direction, such as the root-mean-square
speed in thermal equilibrium at a given temperature. However, it is expected
that for a given ground speed, the earthbound particles moving eastward will
have the highest speed with respect to an ECI frame and hence have the
shortest wavelength, while those moving westward, the longest. Thus the
matter wavelength is expected to possess an east-west directional anisotropy
due to earth's rotation, aside from the speed-dependence in the mass. In
order to acquire a matter wavelength of one angstrom (the order of lattice
spacing), the speed for electrons is $7.4\times 10^{6}$ m/sec. This speed is
much higher than the linear speed due to earth's rotation. Obviously, it
makes no substantial difference whether the speed is referred to an ECI
frame or to the ground. Therefore, the local-ether model is actually in
accord with the Davisson-Germer experiment and others dealing with electron
wave, as examined within the present precision.

On the other hand, for heavier particles, such as neutron and atoms, the
effects of earth's rotation would be more appreciable. For example, in order
to acquire a matter wavelength of one angstrom, the speed for neutrons is $%
4\times 10^{3}$ m/sec, which is higher than the linear speed due to earth's
rotation merely by a factor of about 10. In the Bragg reflection by a
geostationary crystal, the ground speed of the reflected particles is
identical to that of the incident ones. However, the wavelength of the
reflected matter wave is expected to be different from that of the incident
one, except when the Bragg angle\ is very small as measured from the crystal
plane. Moreover, the Bragg angle and the fringe period in Young's slit
diffraction are expected to depend on the orientation of the experimental
setup with respect to the ground, even when the ground speed of the incident
particles is identical.

$\ $

\noindent {\bf 5.2.\ Predicted anisotropy in Bragg angle due to earth's
rotation}

We then proceed to examine the minute effect due to the laboratory velocity.
Suppose the Bragg angle $\theta _{B}$ is measured from the lattice plane
(represented by the vector ${\bf t}$) to the reflected beam. Then the
deflection angle of the particle beam measured from incidence to reflection
is $2\theta _{B}$. For a given incident beam with an incident angle $\theta
^{\prime }$, there are two values for the Bragg angle $\theta _{B}$, one
positive and one negative, which are associated with two orientations of the
crystal symmetric with respect to the incident beam, where $\theta _{B}=\pm
(\pi /2-\theta ^{\prime })$ and $0<\theta ^{\prime }<\pi /2$. Consider the
case where the direction of the laboratory velocity ${\bf v}_{0}$ lies in
the plane of incidence. Suppose the incident beam makes an angle $\alpha $
as measured from the direction of ${\bf v}_{0}$. Thus the vector ${\bf t}$
makes an angle $\alpha +\theta _{B}$ from that direction. Then the Bragg
condition is given by 
$$
m_{0}[v_{p}+v_{0}\cos (\alpha +\theta _{B})\cos \theta _{B}]d|\sin \theta
_{B}|=n\pi \hbar .\eqno
(23) 
$$
It is seen that the laboratory velocity complicates the dependence on angle $%
\theta _{B}$.

Consider the Bragg reflection of neutron wave from a pyrolytic graphite
crystal briefly reported in [15], where the laboratory is geostationary and
the incident neutron beam is directed due south ($\alpha =90^{\circ }$).
Then the Bragg condition becomes 
$$
|\sin \theta _{B}|\left\{ 1-(v_{0}/v_{p})\sin \theta _{B}\cos \theta
_{B}\right\} =n\pi \hbar /m_{0}v_{p}d.\eqno
(24) 
$$
The lattice constant is reported to be $d=6.708$ angstrom with $n=4$; the
neutron wavelength is estimated to be $\lambda =1.134$ angstrom and hence it
is understood that $v_{p}=3490$ m/sec; and the laboratory is located at a
latitude of $38.63^{\circ }$ and hence $v_{0}=362$ m/sec [15]. Then,
according to the preceding formula, the predicted Bragg angles are $\theta
_{a}=20.48^{\circ }$ and $\theta _{b}=-19.11^{\circ }$. A positive or
negative Bragg angle corresponds to the situation that the beam is reflected
toward west or east, respectively. It is noted that the beam reflected
toward west deflects more in magnitude than the one toward east. The
difference in magnitude between the two Bragg angles is associated with the
speed ratio $v_{0}/v_{p}$ and presents an anisotropy due to earth's rotation.

The constructive reflection at the Bragg angle results in a minimum
transmission which in turn can be detected by a counter. The measured
results of the crystal rotation angle for minimum transmission are $\theta
_{1}=19.14^{\circ }$ and $\theta _{2}=159.63^{\circ }$ [15], where the
orientation corresponding to the starting value of the crystal rotation
angle is not specified explicitly. It is understood to correspond to the
orientation at which the normal of the reflecting crystal surface is
directed due west. Thus $\theta _{1}$ and $\theta _{3}$ ($=\theta
_{2}-180^{\circ }=-20.37^{\circ }$) should be identical to $\theta _{a}$ and 
$\theta _{b}$, respectively. It is seen that the two Bragg reflections are
asymmetric, as predicted by the local-ether model. However, the agreement is
not good quantitatively. The discrepancies in the Bragg angles might be due
to that the incident beam is not exactly directed due south or that the
crystal rotation angle should be understood otherwise. Anyway, the predicted
directional anisotropy in the Bragg angle due to earth's rotation provides a
means to test the local-ether wave equation.

$\ $

\noindent {\bf 5.3. Matter-wave Sagnac effect in rotating loop}

By using the Bragg reflection from multiple crystals to form a closed path
for the particle beam, the loop interferometry of matter wave can be
achieved. We discuss the Sagnac effect of matter wave associated with the
rotation of an interferometer which has a closed path formed by a series of
reflecting planes of suitable orientations. Consider two coherently split
particle beams form two paths $L_{1}$ and $L_{2}$ which in turn form a
coplanar closed contour $L$ of arbitrary shape.{\rm \ }Suppose the loop $L$
is rotating about an axis at an arbitrary location with an arbitrary
directed rotation rate $\bar{\omega}_{I}$ with respect to the laboratory
frame which in turn is rotating in an ECI frame. Thus the velocity ${\bf v}%
_{l}$ of the various reflecting planes forming the path is not uniform and
is given by ${\bf v}_{l}={\bf v}_{0}+(\bar{\omega}_{I}+\bar{\omega}%
_{E})\times ({\bf r}-{\bf r}_{0})$ with respect to the local-ether frame,
where{\rm \ }${\bf r}$ and ${\bf r}_{0}$ denote the position vectors of the
planes and of a suitable reference point at which the laboratory velocity $%
{\bf v}_{0}$ is defined, respectively, and $\bar{\omega}_{E}$ is the
directed rate of earth's rotation.

The direction of the particle beam around the loop is determined by the
propagation vector ${\bf k}_{p}$ ($=m{\bf v}_{p}/\hbar $), where ${\bf v}%
_{p} $ ($={\bf v}-{\bf v}_{l}$) is the particle velocity with respect to the
associated reflecting plane. For the trivial case with $\omega _{I}=0$, the
path velocity ${\bf v}_{l}$ is substantially uniform among the planes, as $%
\omega _{E}$ is quite small. Thus the particle velocity ${\bf v}_{p}$, the
propagation vector ${\bf k}_{p}$, and the particle beam traveling between
two consecutive planes are all along the directed linear segment ${\bf l}$
joining the reflecting point on one of the two planes to that on the other.
Consequently, the term ${\bf k}_{p}\cdot {\bf l}$ incorporated in the phase
variation ${\bf k\cdot l}$ over the path segment ${\bf l}$ is then equal to
the scalar product $k_{p}l$. Furthermore, the particle speed $v_{p}$ and the
propagation constant $k_{p}$ are always fixed while the particle beam is
traveling around the loop. On the other hand, for the case with $\omega
_{I}\neq 0$, the path velocity ${\bf v}_{l}$ is varying among the planes.
Thus the direction of the particle beam, as well as those of the particle
velocity ${\bf v}_{p}$ and of the propagation vector ${\bf k}_{p}$, tends to
deviate from the path segment ${\bf l}$. However, the effect of these
deviations on the phase variation is small if the rotation rate $\omega _{I}$
times the segment length $l$ is much smaller than the particle speed $v_{p}$%
. Actually, it is merely of the second order of the speed ratio $\omega
_{I}l/v_{p}$, since the velocity difference between two consecutive planes
under rotation is perpendicular to the segment ${\bf l}$. Similarly, the
fractional variation in the magnitude $k_{p}$ around the loop is also of
this order.

Thereby, for a loop formed by closely spaced reflecting planes or undergoing
a slow rotation such that the condition $\omega _{I}l\ll v_{p}$ is met for
every segment, the phase difference between the coherent beams along the two
paths can be given by the path integral 
$$
\triangle \phi =-\int_{L_{1}}{\bf k}\cdot \hat{l}_{1}dl\ +\ \int_{L_{2}}{\bf %
k}\cdot \hat{l}_{2}dl=\oint_{L}{\bf k}\cdot d{\bf l},\eqno
(25)
$$
where $d{\bf l}=-\hat{l}_{1}dl$ and $\hat{l}_{2}dl$ on paths $L_{1}$ and $%
L_{2}$, respectively. Further, as the differential phase variation ${\bf k}%
_{p}{\bf \cdot }d{\bf l}$ can be given by the scalar product $k_{p}dl$ in
conjunction with $k_{p}$ being a constant value around the loop, the phase
difference becomes 
$$
\triangle \phi =k_{p}(l_{2}-l_{1})+\frac{m_{0}}{\hbar }\oint_{L}{\bf v}%
_{l}\cdot d{\bf l,}\eqno
(26)
$$
where $l_{1}$ and $l_{2}$ are the lengths of paths $L_{1}$ and $L_{2}$,
respectively. In deriving the preceding formula, the ordinary condition $%
v\ll c$ is assumed such that the speed-dependent mass $m$ is substantially
identical to the rest mass $m_{0}$.

For the case where the path is suitably structured such that $l_{1}=l_{2}$,
the phase difference then takes the form of 
$$
\triangle \phi =\frac{2\omega _{0}}{c^{2}}(\bar{\omega}_{I}+\bar{\omega}%
_{E})\cdot {\bf S,}\eqno
(27) 
$$
where ${\bf S}$ ($=\frac{1}{2}\oint_{L}{\bf r}\times d{\bf l}$) is the
directed area enclosed by the loop $L$ and we have made use of a vector
identity, the relation $\hbar \omega _{0}=m_{0}c^{2}$, and of the result
that the integration of a constant vector ${\bf v}_{0}$ or ${\bf r}_{0}$
around an arbitrary loop is zero. It is seen that earth's rotation as well
as the rotation of the loop with respect to the ground contributes to the
phase difference, as the path velocity ${\bf v}_{l}$ is referred to an ECI
frame for terrestrial experiments and hence incorporates the linear velocity
due to earth's rotation. The phase-difference formula (27) with either $\bar{%
\omega}_{I}$ or $\bar{\omega}_{E}$ looks like the matter-wave Sagnac effect
derived in [15, 20-22] from various different approaches. The matter-wave
Sagnac effect associated with the rotation-induced quantum interference has
been demonstrated with a particle beam of electron [22], neutron [15, 16],
or of atom [23, 24]. In these experiments, the loop can be rotating on a
turntable [22, 23] or simply be geostationary with variable orientation [15,
16]. For a geostationary loop with a specific structure, the phase
difference due to earth's rotation together with that due to earth's gravity
has also been derived from the local-ether model [25], which agrees with the
experimental results with neutron beam reported in [15, 16].

It is noted that the phase difference (27) is identical to the one for
optical wave when the natural frequency $\omega _{0}$ is replaced with the
angular frequency of an optical wave, aside from a factor of 2 due to a
difference in the path structure [3]. As the natural frequency of matter
wave is much higher than an optical frequency, the loop area required for
the matter-wave interference can be much smaller. By using the geostationary
loop interferometer with neutron wave, the phase difference (27) due to
earth's rotation alone has been demonstrated with a high precision. It is
noticed that the loop area employed for the detection of earth's rotation
with neutron wave is about $10^{-3}$ m$^{2}$ in [15, 16], which is much
smaller than $2\times 10^{5}$ m$^{2}$ in the Michelson-Gale experiment for
the same purpose with optical wave [26]. By using cesium atomic beam,
earth's rotation has also been detected with the Sagnac loop interferometry
[24].

Meanwhile, in as early as 1904 Michelson supposed that the Sagnac effect due
to the orbital motion of the Earth around the Sun might be detectable [26],
although the angular speed of the orbital motion is about 1/365 times that
of the rotation. If this orbital effect does exist, it could be demonstrated
with matter wave of neutron or atom by constructing a path of area of $0.2$ m%
$^{2}$ or smaller. However, the orbital effect is never observed in
terrestrial interferometers either with electromagnetic or matter wave, to
our knowledge. In some earthbound experiments with electromagnetic wave,
this effect has been assumed to be null by resorting to the principle of
local Lorentz invariance (see [3] for a discussion). Anyway, according to
the local-ether propagation model, the particle velocity which determines
the propagation vector ${\bf k}$ of matter wave is referred uniquely to an
ECI frame, rather than to a heliocentric inertial, the ECEF (earth-centered
earth-fixed), or any other frame. Thus earthbound experiments can depend on
earth's rotation but are entirely independent of earth's orbital motion
around the Sun or whatever. Thereby, the local-ether model unambiguously
predicts a discrepancy between the effects of the rotational and the orbital
motions of the Earth in the Sagnac loop interferometry as well as in many
other earthbound experiments discussed in [3, 4], which provides a means to
test its validity. Moreover, the phase-difference formula (27) is not
expected to hold for an interferometer with a high rotation rate, especially
when the particle speed $v_{p}$ is low. This restriction provides another
means to test the local-ether wave equation.

$\ $

\noindent {\large {\bf 6. Conclusion}}

The local-ether wave equation for a free particle looks like the free-space
Klein-Gordan equation, except the reference frame of the time derivative. By
virtue of the natural frequency, the wave equation leads to a dispersion
relation for a harmonic-like wave packet. From the dispersion relation, the
angular frequency and wavelength of matter wave and the speed-dependent mass
of particle are derived, which look like the postulates of de Broglie and
the Lorentz mass-variation law, respectively, except the reference frame of
the particle velocity. For a harmonic matter wave, the angular frequency and
the propagation vector are associated with the energy and the momentum of
the particle, respectively, as the energy and the momentum are given by the
expectation values of the temporal and the spatial derivatives evaluated in
the wavefunction, respectively. As the phase variation of a particle beam is
given by the propagation vector and the path length along the beam, a shift
in these quantities leads to a phase difference between two coherent beams.
From the difference in the path length, the Bragg condition of constructive
reflection from a crystal is derived. Moreover, from the shift in the
propagation vector due to earth's rotation, the rotation of the
interferometer, or to the gravity, the phase differences in the loop
interferometry are derived.

As the velocity of earthbound particles is referred uniquely to an ECI
frame, rather than a laboratory frame or any other, it is predicted that
earth's rotation leads to an east-west directional anisotropy both in mass
and wavelength. Moreover, the particle velocity involved in the matter-wave
interferometry incorporates the laboratory velocity with respect to an ECI
frame for terrestrial experiments. For electrons, the mass is light and the
speed is normally high. The anisotropy due to earth's rotation or the effect
due to the laboratory velocity is then negligibly small. Thereby, the
local-ether model for electron wave is actually in accord with the
Davisson-Germer experiment and Young's slit diffraction as examined within
the present precision. Further, the effect of the laboratory velocity can
cancel out in a loop interferometer. Consequently, in spite of the
restriction on the reference frame of particle velocity, the derived
interference formulas for the Bragg reflection or for the loop
interferometry can be independent of the laboratory velocity and then comply
with Galilean relativity.

On the other hand, for heavier particles of neutron or atom, the effects of
earth's rotation on matter wavelength are expected to be more appreciable.
It is predicted that the Bragg angle and the fringe period in Young's slit
diffraction depend on the orientation of the experimental setup with respect
to the ground. For a given incident neutron beam, it is analyzed that the
Bragg angle can deviate by an amount of about one degree between two
symmetric orientations of the crystal. The predicted directional anisotropy
in the Bragg angle and in the fringe period, the slow-rotation restriction
on the famous formula of matter-wave Sagnac effect, and the discrepancy
between the effects of earth's rotational and orbital motions provide
different approaches to test the local-ether wave equation.

$\ $

$\ $

\noindent {\large {\bf References}}

\begin{itemize}
\item[{\lbrack 1]}]  See, for example, {R.M. Eisberg, {\it Fundamentals of
Modern Physics}} ({Wiley, New York, 1961), chs. 3 and 6.}

\item[{\lbrack 2]}]  H.A. Lorentz, in {\it The Principle of Relativity}
(Dover, New York, 1952), p. 11.

\item[{\lbrack 3]}]  C.C. Su, {\it Eur. Phys. J. C} {\bf 21}, 701 (2001); 
{\it Europhys. Lett}. {\bf 56}, 170 (2001).

\item[{\lbrack 4]}]  C.C. Su, {\it Eur. Phys. J. B} {\bf 24}, 231 (2001).

\item[{\lbrack 5]}]  C.C. Su, {\it J. Electromagnetic Waves Applicat.} {\bf %
16}, 403 (2002).

\item[{\lbrack 6]}]  C.C. Su, ``A local-ether wave\ equation and the
consequent electromagnetic force law,'' {\it J. Electromagnetic Waves
Applicat.} (in press); in {{\it IEEE Antennas Propagat. Soc. Int}. {\it Symp}%
. {\it Dig.}} (2001), vol. 1, p. 216; in{\ {\it Bull. Am. Phys. Soc.}} (Mar.
2{001}){\it ,} p. 1144{.}

\item[{\lbrack 7]}]  J.D. Bjorken and S.D. Drell, {\it Relativistic Qua}{\it %
ntum Mechanics} (McGraw-Hill, New York, 1964); {{A.A. Sokolov, I.M. Ternov,
and V. Ch. Zhukovskii, {\it Quantum Mechanics}}} ({{Mir Publishers, Moscow,
1984), sect. 17; F. Gross, {\it Relativistic Quantum Mechanics and Field
Theory}}} (Wiley, New York, 1993), ch. 4.

\item[{\lbrack 8]}]  H. Kragh, {\it Am. J. Phys}. {\bf 52}, 1024 (1984).

\item[{\lbrack 9]}]  E.P. Battey-Pratt and T.J. Racey, {\it Int. J. Theoret.
Phys}. {\bf 19}, 437 (1980). (In this paper, the mass of a particle is
ascribed to the spin.)

\item[{\lbrack 10]}]  See, for example, S. Ramo, J.R. Whinnery, and T. Van
Duzer, {\it Fields and Waves in Communication Electronics} (Wiley, New York,
1965), pp. 341 and 384.

\item[{\lbrack 11]}]  {J.D. Jackson, {\it Classical Electrodynamics} (Wiley,
New York, 1975), ch. 11.}

\item[{\lbrack 12]}]  {P. Lorrain and D.R. Corson, {\it Electromagnetic
Fields and Waves}} ({Freeman, San Francisco, 1972), ch. 5.}

\item[{\lbrack 13]}]  L.C. Baird, {\it Am. J. Phys}. {\bf 48}, 779 (1980);
P.D. Gupta, {\it Am. J. Phys}. {\bf 49}, 890 (1981).

\item[{\lbrack 14]}]  L.B. Okun, {\it Phys. Today} {\bf 42}, 31 (1989).

\item[{\lbrack 15]}]  J.-L. Staudenmann, S.A. Werner, R. Colella, and A.W.
Overhauser, {{\it Phys. Rev. A} {\bf 21}, 1419 (1980).}

\item[{\lbrack 16]}]  K.C. Littrell, B.E. Allman, and S.A. Werner, {{\it %
Phys. Rev. A} {\bf 56}, 1767 (1997).}

\item[{\lbrack 17]}]  C. J\"{o}nsson, {\it Am. J. Phys}. {\bf 42}, 4 (1974).

\item[{\lbrack 18]}]  A. Zeilinger, R. G\"{a}hler, C.G. Shull, W. Treimer,
and W. Mampe, {\it Rev. Mod. Phys}. {\bf 60}, 1067 (1988).

\item[{\lbrack 19]}]  O. Carnal and J. Mlynek, {\it Phys. Rev. Lett}. {\bf 66%
}, 2689 (1991).

\item[{\lbrack 20]}]  M. Dresden and C.N. Yang, {\it Phys. Rev. D} {\bf 20},
1846 (1979).

\item[{\lbrack 21]}]  L.A. Page, {\it Phys. Rev. Lett}. {\bf 35}, 543 (1975).

\item[{\lbrack 22]}]  {F. Hasselbach and M. Nicklaus, {\it Phys. Rev. A} 
{\bf 49}, 143 (1993).}

\item[{\lbrack 23]}]  A. Lenef, T.D. Hammond, E.T. Smith, M.S. Chapman, R.A.
Rubenstein, and D.E. Pritchard, {\it Phys. Rev. Lett}. {\bf 78}, 760 (1997).

\item[{\lbrack 24]}]  T.L. Gustavson, P. Bouyer, and M.A. Kasevich, {\it %
Phys. Rev. Lett}. {\bf 78}, 2046 (1997).

\item[{\lbrack 25]}]  C.C. Su, ``Reinterpretation of the effects of earth's
rotation and gravity on the neutron-interferometry experiment,'' {\it %
Europhys. Lett}. (in press).

\item[{\lbrack 26]}]  R. Anderson, H.R. Bilger, and G.E. Stedman, {\it Am.
J. Phys}. {\bf 62}, 975 (1994).
\end{itemize}

\end{document}